\documentstyle[12pt]{article}
\begin{document}

\begin{center}
\null\vspace{2cm}
{\large {\bf Hawking Radiation of Schwarzschild-de Sitter Black Hole by Hamilton-Jacobi method}}\\
\vspace{2cm}
M. Atiqur Rahman\footnote{E-mail: $atirubd@yahoo.com$}\\
{\it Department of Applied Mathematics, Rajshahi University}\\
M. Ilias Hossain\footnote{E-mail: $ilias_-math@yahoo.com$}\\
{\it Department of Mathematics, Rajshahi University}\\
 {\it Rajshahi - 6205, Bangladesh}

\end{center}
\vspace{3cm}
\centerline{\bf Abstract}
\baselineskip=18pt
\bigskip

We investigate the Hawking radiation of Schwarzschild-de Sitter (SdS) black hole by massive particles tunneling  method. We consider the spacetime background to be dynamical, incorporate the self-gravitation effect of the emitted particles and show that the tunneling rate is related to the change of Bekenstein-Hawking entropy and the derived emission spectrum deviates from the pure thermal spectrum when energy and angular momentum are conserved. Our result is also in accordance with Parikh and Wilczek\rq s opinion and gives a correction to the Hawking radiation of SdS black hole.
\vspace{0.5cm}
{\it PACS number(s)}: 04.70.-s, 04.70.Dy, 97.60.Lf
\vfill

\newpage

\section{Introduction}\label{sec1}
Hawking radiation is viewed as tunneling process caused by vacuum fluctuation near the event horizon of black hole \cite{one,two}. A method to describe Hawking radiation as tunneling process was first developed by Kraus and Wilczek \cite{three,four} and then reinterpreted  by Parikh and Wilczek \cite{five} as quantum tunneling by considering a particle with negative energy just inside, a positive energy just outside the horizon which can be explained as a virtual particle pair spontaneously created near the horizon of black hole and materializes as a true particle. The particle with negative energy tunnels into the horizon and is absorbed, while the particle with positive energy left outside the horizon to infinite distance and forms the Hawking radiation.

From the past decade the tunneling method has been successfully applied to deal with Hawking radiation of black holes. A lot of works for various spacetimes \cite{six,seven,eight,nine,ten,eleven,twelve,thirteen,fourteen,fifteen,sixteen,seventeen,eighteen,nineteen,twenty,twenty one,twenty two,twenty three,twenty four,twenty five,twenty six,twenty seven,twenty eight,twenty nine,thirty,thirty one,thirty two} show its validity and all of these are limited to massless particle. Based on the above tunneling picture, two different methods have been employed to calculate the imaginary part of the action, one by Parikh and Wilczek \cite{three,four} and other by Angheben et al. \cite{thirty three} named as null-geodesic and Hamilton-Jacobi methods respectively. In fact, the method of Angheben et al. \cite{thirty three} is an extension of the complex path analysis proposed by Padmanabhan et al. \cite{thirty four, thirty five, thirty six}. On the other hand, Hawking radiation from massive uncharged particle tunneling \cite{thirty seven} and charged particle tunneling \cite{thirty eight} were proposed by Zhang and Zhao. Following this work, few researches have been carried out as charged particle tunneling \cite{thirty nine,fourty,fourty one,fourty two}.

Recently, Kerner and Mann developed quantum tunneling methods for analyzing the temperature of Taub-NUT black holes \cite{fourty three} using both the null-geodesic and Hamilton-Jacobi methods. The latter method involve calculating the relativistic Hamilton-Jacobi equation in which the drive radiation spectrum was only a leading term due to the fact that the self-gravitation interaction and energy conservation of emitted particle were ignored. According to the Parikh and Wilczek opinion the true radiation spectrum is not strictly thermal but satisfies the underlying unitary theory when self-gravitation interaction and energy conservation are considered. It is clear that the background geometry of a radiating black hole should be altered (unfixed) with the loss of energy. Taking the self-gravitation interaction and unfixed background spacetime into account Chen, Zu and Yang reformed Hamilton-Jacobi method for massive particle tunneling and investigate the Hawking radiation of the Taub-NUT black hole \cite{fourty four}. Connecting this method Hawking radiation of Kerr-NUT black hole \cite{fourty five} and the charged black hole with a global monopole \cite{fourty six} have been developed. We apply these method to investigate the Hawking radiation of SdS black hole. Since our prime concern of this work is to calculate the imaginary part of action from Hamilton-Jacobi equation avoid by exploring the equation of motion of the radiation particle in Painlev\'e coordinate system and calculating the Hamilton equation. We need not differentiate radiation particle, although the equation of motion of massive particle is different from massless particle. After considering the self-gravitational interaction and the unfixed background spacetime, the derived radiation spectrum deviates from the purely thermal one and the tunneling rate is related to the change of Bekenstein-Hawking entropy.

Study of Hawking radiation on black holes with a positive cosmological constant become important due to the two reasons. one, the recent observed accelerating expansion of our universe indicates the cosmological constant might be a positive one \cite{fourty seven,fourty eight,fourty nine}, and conjecture about de Sitter/CFT correspondence \cite{fifty,fifty one}. For black hole with positive cosmological constant particles can be created at both black hole and cosmological horizon and there exists different tunneling behaviors. The outgoing and incoming particles tunnel from black hole and cosmological horizon respectively and formed Hawking radiation. For black hole horizon, the incoming particles can fall into the horizon along classically permitted trajectories but for cosmological horizon outgoing particles can fall classically out of the horizon. So our study of black hole in de Sitter space is important and meaningful.

This paper is organized as follows. In section 2, we describe the SdS black hole spacetime with the position of event horizon. Near the event horizon the new line element of SdS black hole is also derived here. Taking the unfixed background spacetime and the self-gravitational interaction into account, we review the Hawking radiation of SdS black hole from massive particle tunneling method in section 3. Finally, in section 4, we present our remarks.

\section{Schwarzschild-de Sitter black hole}\label{sec2}
The Schwarzschild-de Sitter black hole, which is the solution
of Einstein equations with a positive $\Lambda(=3/\ell^2)$ term
corresponding to a vacuum state spherical symmetric configuration of the
form
\begin{eqnarray}
ds^2&=&g_{\mu \nu }dx^\mu dx^\nu\nonumber\\
 &=&-\left(1-\frac{2m}{r}-\frac{r^2}{\ell^2}\right)dt^2
 +\left(1-\frac{2m}{r}-\frac{r^2}{\ell^2}\right)^{-1}dr^2+r^2(d\theta^2+\textrm{sin}^2\theta d\varphi^2),\label{eq1}
\end{eqnarray}
where $m$ being the mass of the black hole and the coordinates are defined
such that $-\infty\leq t\leq \infty $, $r\geq 0$, $0\leq \theta \leq
\pi $, and $0\leq \phi \leq 2\pi$. At large $r$, the metric
(\ref{eq1}) tends to the dS space limit. The explicit dS case is
obtained by setting $m=0$ while the explicit Schwarzschild case is
obtained by taking the limit $\ell\rightarrow\infty$. When $\ell^2$
is replace by $-\ell^2$, the metric (\ref{eq1}) describes an
interesting nonrotating AdS black hole called the Schwarzschild
Anti-de Sitter (SAdS) black hole.

The horizons of the SdS black hole are located at the real positive
roots of
$\frac{1}{\ell^2r}(r-r_h)(r-r_c)(r_--r)=0$, and
there are more than one horizon if $0<\Xi<1/27$ where
$\Xi=M^2/\ell^2$. The black hole (event) horizon $r_h$ and the
cosmological horizon $r_c$ are located, respectively, at
\begin{eqnarray}
r_h&=&\frac{2m}{\sqrt{3\Xi}}\textrm{cos}\frac{\pi+\psi}{3}\label{eq2},\\
r_c&=&\frac{2m}{\sqrt{3\Xi}}\textrm{cos}\frac{\pi-\psi}{3}\label{eq3},
\end{eqnarray}
where
\begin{equation}
\psi=\textrm{cos}^{-1}(3\sqrt{3\Xi})\label{eq4}.
\end{equation}
In the limit $\Xi\rightarrow0$, one finds
that $r_h\rightarrow 2m$ and $r_c\rightarrow \ell$, and it is
obvious that $r_c>r_h$, i.e., the event horizon is the smallest
positive root. The spacetime is dynamic for $r<r_h$ and $r>r_c$. The
two horizons coincide: $r_h=r_c=3m$ (extremal), when $\Xi=1/27$, and
the spacetime then becomes the well known Nariai spacetime.
Expanding $r_h$ in terms of $m$ with $\Xi<1/27$, we obtain
\begin{equation}
r_h=2m\left(1+\frac{4m^2}{\ell^2}+\cdot
\cdot\cdot\right)\label{eq5},
\end{equation}
that is, the event horizon of the SdS black hole is greater than the
Schwarzschild event horizon, $r_{Sch}=2m$. For $\Xi>1/27$, the
spacetime is dynamic for all $r>0$, that is, the metric (\ref{eq1})
then represents not a black hole but an unphysical naked singularity
at $r=0$.
For the convenient of discussion, we define $\Delta=r^2-2mr-\frac{r^4}{\ell^2}$ and then the line element becomes
\begin{equation}
ds^2=-\frac{\Delta}{r^2}dt^2+\frac{r^2}{\Delta}dr^2+r^2(d\theta^2+{\rm sin^\theta}d\phi^2)\label{eq6}.
\end{equation}
The position of black hole horizon is same as given in Eq. (\ref{eq5}).
Near the black hole horizon, the line element takes of the form
\begin{equation}
ds^2=-\frac{\Delta_{,r}(r_h)(r-r_h)}{r^2_h}dt^2+\frac{r^2_h}{\Delta_{,r}(r_h)(r-r_h)}dr^2+r^2_h(d\theta^2+{\rm sin^2\theta}d\phi^2)\label{eq7},
\end{equation}
where,
\begin{equation}
\Delta_{,r}(r_h)=\frac{d\Delta}{dr}\bigg |_{r=r_h}=2(r_h-m-2\frac{r^3_h}{\ell^2}).\label{eq8}
\end{equation}
Since the event horizon of SdS black hole coincides with the outer infinite redshift surface, here we can apply the geometrical optics limit. Within WKB approximation \cite{fifty two} the relationship between the tunneling rate and the action of the radiative particle is as
\begin{eqnarray*}
\Gamma \sim {\rm exp}(-2{\rm Im}I).
\end{eqnarray*}
\section{The Hamilton-Jacobi Method}\label{sec4}
Here we used the method of Chen et al. \cite{fourty four} to discuss the Hawking-Radiation from the action of radiation particles. As mention before this method is different from Parikh and Wilczek method in which the action mainly relies on the exploration of the equation of motion in the Painlev\'e coordinates systems and the calculation of Hamilton equation. In the Hamilton-Jacobi method we avoid this and calculate the imaginary part of the action from the relativistic Hamilton-Jacobi equation.

The action $I$ of the outgoing particle from the black hole horizon satisfies the relativistic Hamilton-Jacobi equation
\begin{equation}
g^{\mu\nu}\left(\frac{\partial I}{\partial x^\mu}\right)\left(\frac{\partial I}{\partial x^\nu}\right)+u^2=0,\label{eq9}
\end{equation}
in which $u$ and $g^{\mu\nu}$ are the mass of the particle and the inverse metric tensors derived from the line element (\ref{eq7}).

For the metric (\ref{eq7}), we get non-null inverse metric tensors
\begin{eqnarray}
g^{00}=-\frac{r^2_h}{\Delta_{,r}(r_h)(r-r_h)}, \quad g^{11}=\frac{\Delta_{,r}(r_h)(r-r_h)}{r^2_h}, \quad g^{22}=\frac{1}{r_h^2}, \quad g^{33}=\frac{1}{r_h^2{\rm sin^2\theta}}.\label{eq10}
\end{eqnarray}
Using Eq. (\ref{eq10}), we have from Eq. (\ref{eq9})
\begin{equation}
-\frac{r^2_h}{\Delta_{,r}(r_h)(r-r_h)}\left(\frac{\partial I}{\partial t}\right)^2+\frac{\Delta_{,r}(r_h)(r-r_h)}{r^2_h}\left(\frac{\partial I}{\partial r}\right)^2+\frac{1}{r_h^2}\left(\frac{\partial I}{\partial \theta}\right)^2+\frac{1}{r_h^2{\rm sin^2\theta}}\left(\frac{\partial I}{\partial \phi}\right)^2+u^2=0\label{eq11}.
\end{equation}
It is very difficult to solve the action $I$ for $I(t, r, \theta, \phi)$. Considering the properties of black hole spacetime, the separation of variables can be taken as follows
\begin{equation}
I=-\omega t+R(r)+H(\theta)+j\psi\label{eq12},
\end{equation}
where $\omega$ and $j$ are respectively the energy and angular momentum of the particle. Since SdS black hole is nonrotating, the angular velocity of the particle at the horizon is $\Omega_h=\frac{d\varphi}{dt}\Bigg |_{r=r_h}=0$.
Using Eq. (\ref{eq12}) into Eq. (\ref{eq11}) and solving $R(r)$ yields
\begin{eqnarray}
R(r)=\pm\frac{r^2_h}{\Delta_{,r}(r_h)}\int \frac{dr}{(r-r_h)}\quad \times \sqrt{\omega^2-\frac{\Delta_{,r}(r_h)(r-r_h)}{r^2_h}\left[g^{22}(\partial_\theta H(\theta))^2+g^{33}j^2+u^2\right]}\label{eq13}.
\end{eqnarray}
We consider the emitted particle as an ellipsoid shell of energy  to tunnel across the event horizon and should not have  motion in  $\theta$-direction ($d\theta=0$) and therefore, finishing the above integral we get
\begin{eqnarray}
R(r)=\pm \frac{\pi i r^2_h}{\Delta_{,r}(r_h)}\omega+\xi \label{eq14},
\end{eqnarray}
where $\pm$ sign comes from the square root and $\xi$ is the constant of integration. Inserting Eq. (\ref{eq14}) into Eq. (\ref{eq12}), the imaginary part of two different actions corresponding to the outgoing and incoming particles can be written as
\begin{eqnarray}
{\rm Im}I_\pm =\pm \frac{\pi r^2_h}{\Delta_{,r}(r_h)}\omega+{\rm Re}(\xi) \label{eq15}.
\end{eqnarray}
In the classical limit \cite{fifty three}, we ensure the incoming probability to be unity when there is no refection i.e., every thing is absorbed by the horizon. In this situation the appropriate value of $\xi$ instead of zero or infinity can be taken as $\xi=\frac{\pi i r^2_h}{\Delta_{,r}(r_h)}\omega+{\rm Re}(\xi)$. Therefore, ${\rm Im}I_-=0$ and $I_+$ give the imaginary part of action $I$ corresponding to the outgoing particle of the form
\begin{eqnarray}
{\rm Im}I &=& \frac{\pi r^2_h}{\Delta_{,r}(r_h)}\omega \nonumber\\
&=&\frac{\pi r^2_h }{(r_h-m-2\frac{r^3_h}{\ell^2})}\omega\label{eq16}.
\end{eqnarray}
Using Eq. (\ref{eq5}) into Eq. (\ref{eq16}), we get the imaginary part of action as
\begin{eqnarray}
{\rm Im}I=\frac{\pi 4m^2\left(1+\frac{4m^2}{\ell^2}+\cdot
\cdot\cdot\right)^2}{2m\left(1+\frac{4m^2}{\ell^2}+\cdot
\cdot\cdot\right)-m-\frac{2}{\ell^2}\{2m\left(1+\frac{4m^2}{\ell^2}+\cdot
\cdot\cdot\right)\}^3}\omega\label{eq16}\label{eq17}.
\end{eqnarray}
Since the SdS spacetime is dynamic, we fix the Amowitt-Deser-Misner (ADM) mass of the total spacetime and allow the SdS black hole to fluctuate. When a particle with energy $\omega$ tunnels out, the mass of the SdS black hole changed into $m-\omega$. Since the angular velocity of the particle at the horizon is zero $(\Omega_h=0)$, the angular momentum is equal to zero. Taking the self-gravitational interaction into account, the imaginary part of the true action can be calculated from Eq. (\ref{eq16}) in the following integral form
\begin{eqnarray}
{\rm Im}=\pi\int^\omega_0\frac{4m^2\left(1+\frac{4m^2}{\ell^2}+\cdot
\cdot\cdot\right)^2}{2m\left(1+\frac{4m^2}{\ell^2}+\cdot
\cdot\cdot\right)-m-\frac{2}{\ell^2}\{2m\left(1+\frac{4m^2}{\ell^2}+\cdot
\cdot\cdot\right)\}^3}d\omega'\label{eq18}.
\end{eqnarray}
Replacing $m$ by $m-\omega$ we have
\begin{eqnarray}
{\rm Im}I=-\pi\int^{(m-\omega)}_m \frac{4(m-\omega')^2\left(1+\frac{4(m-\omega')^2}{\ell^2}+\cdot
\cdot\right)^2}{2(m-\omega')\left(1+\frac{4(m-\omega')^2}{\ell^2}+\cdot
\cdot\right)-(m-\omega')-\frac{2}{\ell^2}\{2(m-\omega')\left(1+\frac{4(m-\omega')^2}{\ell^2}+\cdot
\cdot\right)\}^3}\nonumber\\
\times d(m-\omega')\label{eq19}.
\end{eqnarray}
Within WKB approximation, we can neglect the terms $(m-\omega')^n$ for $n\ge 5 $. Therefore, we rewrite Eq. (\ref{eq19}) of the form
\begin{eqnarray}
{\rm Im}I&=&-4\pi\int^{(m-\omega)}_m \frac{(m-\omega')\left(1+\frac{8(m-\omega')^2}{\ell^2}\right)}{\left(1-\frac{8(m-\omega')^2}{\ell^2}\right)}\times d(m-\omega'),\nonumber\\
&=&-\frac{\pi}{2}\left[4(m-\omega)^2\left(1+\frac{8(m-\omega)^2}{\ell^2}\right)-4m^2\left(1+\frac{4m^2}{\ell^2}\right)\right]\label{eq20}.
\end{eqnarray}
Therefore, the tunneling rate for SdS black hole is given by
\begin{eqnarray}
\Gamma \sim {\rm exp}(-2{\rm Im}I)&=&{\rm exp}\left\{\pi\left[4(m-\omega)^2\left(1+\frac{8(m-\omega)^2}{\ell^2}\right)-4m^2\left(1+\frac{4m^2}{\ell^2}\right)\right]\right\}\nonumber\\
&=&{\rm exp}[\pi(r^2_f-r^2_i)]\nonumber\\
&=&{\rm exp}(\Delta S_{BH}).\label{eq21}
\end{eqnarray}
Here, $r_f=2(m-\omega)\left(1+\frac{4(m-\omega)^2}{\ell^2}\right)$ and $r_i=2m\left(1+\frac{4m^2}{\ell^2}\right)$
are the locations of the SdS event horizon before and after the particle emission, and $\Delta S_{BH}=S_{BH}(m-\omega)-S_{BH}(m)$ is the change of Bekenstein-Hawking entropy. It is clear from Eq. (\ref{eq21}) that the radiation spectrum is not pure thermal although gives a correction to the Hawking radiation of SdS black hole. Expanding the tunneling rate in power of $\omega$ upto second order, the purely thermal spectrum can be derived from Eq. (\ref{eq21}) as discussed by Liu et al. \cite{fourty five} of the form
\begin{eqnarray}
\Gamma \sim {\rm exp}(\Delta S_{BH})&=&{\rm exp}\left\{-\omega \frac{\partial S_{BH}(m)}{\partial\omega}+\omega^2\frac{\partial^2 S_{BH}(m)}{\partial\omega^2}\right\}\nonumber\\
&=&{\rm exp}\left\{-8\pi \omega \left[\left(m+\frac{16m^3}{\ell^2}\right)-\frac{\omega}{2}\left(1+\frac{48m^2}{\ell^2}\right)\right]\right\}.\label{eq22}
\end{eqnarray}
When $\ell \rightarrow \infty$, the pure thermal spectrum can be reduced for Schwarzschild black hole as $\Gamma \sim {\rm exp}(\Delta S_{BH})={\rm exp}\{-8\pi\omega(m-\frac{\omega}{2})\}$. Obviously, the result is in accordance with the result of Parikh and Wilczek \cite{five}. The radiation spectrum given by Eq. (\ref{eq22}) is more accurate and provides an interesting correction to Hawking pure thermal spectrum.

\section{Concluding Remarks}\label{sec4}
In this paper,  we have presented the Hawking radiation as massive particle tunneling method from SdS black hole. We have found that the tunneling rate at the event horizon of SdS black hole is related to the Bekenstein-Hawking entropy, and the factual radiation spectrum deviates from the precisely thermal one when energy conservation and self-gravitational interaction are taken into account. Specially, when $\ell \rightarrow \infty$, i.e., $\Lambda =0$,  the SdS black hole reduced to Schwarzschild black hole. The positions of the event horizon of Schwarzschild black hole before and after the emission of the particles with energy $\omega$ are $r_i=2m$ and $r_f=2(m-\omega)$. From Eq. (\ref{eq21}), the tunneling rate of Schwarzschild black hole can be written as
\begin{eqnarray}
\Gamma \sim {\rm exp}(-2{\rm Im}I)={\rm exp}\left\{\pi\left[4(m-\omega)^2-4m^2\right]\right\}={\rm exp}[\pi(r^2_f-r^2_i)]={\rm exp}(\Delta S_{BH}),\label{eq23}
\end{eqnarray}
which is fully consistent with that obtained by Parikh and Wilczek \cite{five}. We shall further extend our study to the other black holes generalized with
cosmological parameter.\\
{\bf Acknowledgement}\\
One of the authors (MAR) thanks the Abdus Salam International Centre for Theoretical Physics (ICTP), Trieste, Italy, for giving opportunity to utilize its e-journals for research purpose.

\end{document}